\documentclass[singlecolumn,epjc3]{svjour3}
\smartqed  
\RequirePackage{graphicx}
\journalname{Eur. Phys. J. C}
\usepackage{amsmath}
\usepackage{array}
\usepackage{graphicx}
\usepackage{dcolumn}
\usepackage{bm}
\usepackage{caption}

\begin{document}

\title{A new model for spherically symmetric charged compact stars of embedding class one}

\author{S.K. Maurya\thanksref{e1,addr1}
\and Y.K. Gupta\thanksref{e2,addr2}
\and Saibal Ray\thanksref{e3,addr3}
\and Debabrata Deb\thanksref{e4,addr4}.}

\thankstext{e1}{e-mail: sunil@unizwa.edu.om}
\thankstext{e2}{e-mail: kumar$001947$@gmail.com}
\thankstext{e3}{e-mail: saibal@associates.iucaa.in}
\thankstext{e4}{e-mail: d.deb32@gmail.com}

\institute{Department of Mathematical and Physical Sciences,
College of Arts and Science, University of Nizwa, Nizwa, Sultanate
of Oman\label{addr1} \and Department of Mathematics, Raj Kumar
Goel Institute of Technology, Ghaziabad, U.P., India\label{addr2}
\and Department of Physics, Government College of Engineering and
Ceramic Technology, Kolkata 700010, West Bengal,
India\label{addr3} \and Department of Physics, Indian Institute of
Engineering Science and Technology, Shibpur, Howrah 711103, West
Bengal, India\label{addr4}}

\date{Received: date / Accepted: date}

\maketitle

\begin{abstract}
In the present study we search for a new stellar model with
spherically symmetric matter and charged distribution under
general relativistic framework. The model represents a compact
star of embedding class one. The solutions obtain here are general
in their nature having the following two features: firstly, the
metric becomes flat and also the expressions for the pressure,
energy density and electric charge become zero in all the cases if
we consider the constant $A=0$, which shows that our solutions
represent the so-called `electromagnetic mass
model'~\cite{Lorentz1904}, and secondly, the metric function
$\nu(r)$, for the limit $n$ tends to infinity, converts to
$\nu(r)=C{r}^{2}+ ln~B$, which is the same as considered by Maurya
et al.~\cite{Maurya2015a}. We have investigated several physical
aspects of the model and find that all the features are acceptable
within the demand of the contemporary theoretical studies and
observational evidences.
\end{abstract}

\keywords{general relativity, embedding class one, perfect fluid,
electromagnetic field}

\section{Introduction}
It is well known that the $V_n$, which is $n$ dimensional
manifold, can always be embedded in $m[= n(n+1)/2]$ dimensional
Pseudo-Euclidean space. In essence $(m-n)[= n(n-1)/2]$ is the
minimum extra dimensions of pseudo-Euclidean space which is needed
for the embedding class of $V_n$. The embedding class turns out to
be 6 as the relativistic spacetime is 4 dimensional. The class of
general spherical and plane symmetric spacetime are 2 and 3
respectively. The Friedman-Robertson-Lema{\^i}tre
spacetime~\cite{Friedmann1922,Robertson1933,Lemaitre1933} is of
class 1, but the Schwarzschild exterior and interior solutions are
of class 2 and class 1 respectively. Moreover the Kerr metric is
of class 5~\cite{Kuzeev1980}. However, in the present
investigation our discussion is limited to the static spherically
symmetric metric in curvature coordinates which is embedable in
$5-D$ pseudo Euclidean space and hence is of embedding class one
metric.

It is well known that the aforesaid metric is only compatible with
two perfect fluid distributions: the first one is the
Schwarzschild interior solution~\cite{Schwarzschild1916} with de
Sitter's and Einstein's universe as its particular cases, and the
second one is the Kohler-Chao solution~\cite{Kohlar1965}. It is
worth to point out that the former one is conformally flat while
the latter one is non-conformally flat.

Presently we would like to utilize the embedding class one metric
to construct electromagnetic mass models by obtaining charged
perfect fluid distributions. Normally when charge can be made zero
in a charge fluid distribution, then the subsequent distribution
is neutral counterpart of the charged distribution. For example,
if we get Schwarzschild's interior metric after the removal of
charge in a charged fluid, then that is called charged analogue of
the Schwarzschild interior solution~\cite{Schwarzschild1916}. On
the other hand, if the metric of charged fluid turns out to be
flat and also all the physical parameters like pressure, density
vanish, the corresponding charged fluid distribution is said to
form an electromagnetic mass model, i.e. the entire mass is made
up of charge only.

This type of {\it electromagnetic mass model} (EMMM) with
vanishing charge density was first proposed by
Lorentz~\cite{Lorentz1904} and later on by several other
scientists~\cite{Wheeler1962,Feynman1964,Florides1974,Tiwari1984,Gautreau1985,Gron1985,Gron1986,Cooperstock1989,Tiwari1991a,Tiwari1991b,Tiwari1991c,Ray1993,Ray1997,Wilzchek1999,Ray2002a,Ray2002b,Ray2004a,Ray2004b,Ray2006,Ray2007b}.
Unfortunately, in all these electromagnetic mass models the fluid
has negative pressure (tension). However, it is true for gaseous
spheres though at the boundary the vanishing of the density is not
necessary for junction conditions. A model with such a special
type of density have been proposed both for the uncharged as well
as charged cases~\cite{Mehra1980,Kuchowicz1968}. Another idea
about the electromagnetic origin of the electron mass maintains
that, due to vacuum polarization, its interior has the equation of
state of the kind
\begin{equation}
\rho+p=0,
\end{equation}
where $\rho$ and $p$ represents the density and the pressure
respectively. This leads to repulsive pressure  and easier
junction
conditions~\cite{Florides1974,Krori1975,Gron1986,Cooperstock1989}.
It can also be combined with a Weyl-type character of the field
\cite{Ray1993}. The experimental evidence that the electron's
diameter is not larger than $10^{-16}$ cm leads to the conclusion
that the classical model of electron must have a region of
negative density~\cite{Ray1997}.

In this paper we have considered the metric $\nu=n\ln  \left(
1+{{\it Ar}}^{2} \right) +\ln  \left( B \right)$ for $n \geq 2$.
The choice of constraint on $n$ is due to the following reasons:
(i) for $n=0$, there is no meaning of $\nu$ here in the present
context as the spacetime via $\nu$ becomes flat, (ii) for $n=1$,
this reduces to the same as the Kohlar-Chao
solution~\cite{Kohlar1965}, and (iii) for $n < 2$ the term
$(1+Ar^2)^{(n-2)}$ in the expression of $\lambda$ takes the place
in the denominator. We have calculated the data for $n=3.3$ to
$1000$ and wanted to see what would happen in the result for very
high value of $n$. So, one can look in to the Table 2 that if $n$
be large enough i.e. $n=100$, 1000 and even more then $nA$ becomes
approximately a constant, say $nA=C$. This means if we take limit
$n$ tends to infinity then the metric $\nu$ will convert to the
following form $\nu=C{r}^{2}+\ln~B$, which is the same as metric
considered for the solution of EMMM
$(\nu=2\,A{r}^{2}+\ln~B)$~\cite{Maurya2015a}.

In the present work we shall try to form a model for the charged
fluid of class one by assuming specific metric potential(s) of the
class one metric such that they do not form sub set of the metric
potentials of the Kohler-Chao metric~\cite{Kohlar1965} and the
Schwarzschild interior metric (considering de Sitter and Einstein
universe)~\cite{Schwarzschild1916}. Now, if the charge can be made
zero in the charged fluid so obtained, the describing metric will
turn out to be flat by virtue of the class one structure of the
metric.

Outline of the present investigation is as follows: in Sect. 2 the
field equations and some specific results are provided under the
Einstein-Maxwell spacetime whereas we obtain a new class of
solutions in Sect. 3. The matching conditions are discussed in
Sect. 4 and physical properties of the model are explored in Sect.
5. We pass some comments in Sect. 6.

\section{The field equations and the results}

\subsection{The Einstein-Maxwell spacetime}
Let us consider the static spherically symmetric metric in the
form
\begin{equation}
ds^2 = - e^{\lambda(r)} dr^2 - r^2(d\theta^2 + sin^2\theta d\phi^2) + e^{\nu(r)} dt^2.\label{metric1}
\end{equation}

The Einstein-Maxwell field equations can be given as
\begin{equation}
{G^i}_j =  {R^i}_j - \frac{1}{2} R {g^i}_j = \kappa ({T^i}_j +
{E^i}_j),\label{field}
\end{equation}
where $G=1=c$ in relativistic geometric unit and $\kappa=8\pi$ is
the Einstein constant. The matter in the star is expected to be
locally perfect fluid. However ${T^i}_j$ and ${E^i}_j$ are the
energy-momentum tensor of fluid distribution and electromagnetic
field respectively and that can be defined as
\begin{equation}
{T^i}_j = [(\rho + p)v^i\,v_j - p\,{\delta^i}_j],\label{matter}
\end{equation}

\begin{equation}
{E^i}_j = \frac{1}{4\pi}(-F^{im}F_{jm} +
\frac{1}{4}{\delta^i}_jF^{mn}F_{mn}),\label{electric}
\end{equation}
where $\rho$ is the energy density, $p$ is the pressure and $v^i$
is the four-velocity defined as $e^{-\nu(r)/2}v^i={\delta^i}_4$.

\subsection{The embedding class one spacetime}
The metric (\ref{metric1}) may represent spacetime of embedding
class one, if it satisfies the given condition of
Karmarkar~\cite{Karmarkar}
\begin{equation}
R_{{1414}}={\frac
{R_{{1212}}R_{{3434}}+R_{{1224}}R_{{1334}}}{R_{{2323}}}},
\end{equation}
where ${{R_{{2323}}\neq 0}}$~\cite{Pandey1982}.

The above condition with reference to (\ref{metric1}) yields the
following differential equation
\begin{equation}
\label{6}
\frac{\lambda^{\prime}\,e^{\lambda}}{1-\,e^{\lambda}}=-\frac{2\,\nu^{\prime\prime}}{\nu^{\prime}}-\nu^{\prime}.
\end{equation}

The solution of differential Eq. (\ref{6}) can be furnished as
\begin{equation}
\label{elambda} {{\rm e}^{\lambda}}=\left(1+K\frac{{\nu }^{\prime
2}{{\rm e}^{\nu}}}{4}\right),
\end{equation}

\begin{equation}
{\nu}^{\prime }  \left( r \right)\neq 0, \,\,\,{\rm e}
^{\lambda\left( 0 \right)}=1,
\end{equation}
 and
 \begin{equation}
 {\nu}^{\prime }\left( 0 \right)=0,
 \end{equation}
where $K$ is an arbitrary non-zero integration constant.

Using the spherically symmetric metric (\ref{metric1}) and Eq.
(\ref{elambda}), the Einstein-Maxwell field equations can be
written as the following set of equations \cite{Maurya2015a}:
\begin{equation}
8\pi\,p\,=\, \frac
{{\nu}^{\prime}}{{r}^{2}}\left[\frac{4\,r-K{\nu}^{\prime}\,e^{\nu}}{4+K{\nu^{\prime}}^{2}\,e^{\nu}}\right]\,+\,{\frac
{{q}^{2}}{{r}^{4}}},
\end{equation}

\begin{equation}
8\pi\,p\,=\,{\frac {4}{\left(4+K{{\nu}^{\prime}}^{2}{{\rm
e}^{\nu}} \right) }}\,\left[{\frac {{\nu}^{\prime}}{2r}}-{\frac
{\left( K{{\nu}^{\prime}}{{\rm e}^{\nu}}-2\,r \right)\left(
2\,{{\nu}^{\prime\prime}}+{{\nu}^{\prime}}^{2} \right) }{2r \left(
4+K{{\nu}^{\prime}}^{2}{{\rm e}^{\nu}} \right)
}}\right]\,-\,{\frac {{q}^{2}}{{r}^{4}}},
\end{equation}

\begin{equation}
8\pi\,\rho\,=\,{\frac {K{{\rm e}^{\nu}}{{\nu}^{\prime}}}{\left(
4+K{{\nu}^{\prime}}^{2}{ {\rm e}^{\nu}}\right)r }}\,\left[{\frac
{4\left( 2\,{{\nu}^{\prime\prime}}+{{\nu}^{\prime}}^{2} \right)
}{\left( 4+K{{\nu}^{\prime}}^{2}{{\rm e}^{ \nu}} \right) }}+{\frac
{{\nu}^{\prime}}{r}}\right]\, -\,{\frac {{q}^{2}}{{r}^{4}}},
\end{equation}
where the differential with respect to $r$ is denoted by prime.

\section{A new class of solutions}
To determine the expression for the electric charge, we use the
pressure isotropy condition. Therefore, from Eqs. (11) and (12),
we get
\begin{equation}
{\frac {{{q}^{2}}}{{r}^{4}}}\,=\,\left[{\frac
{K{{\nu}^{\prime}}{{\rm e}^{\nu}}}{2r}}-1\right]\,\left[\,\frac
{{\nu}^{\prime}\,(4+K\,{\nu^{\prime}}^{2}{\rm
e}^{\nu})-2\,r\,(2\,{\nu}^{\prime\prime}+{\nu^{\prime}}^{2})}{r\,\left(4+K{\nu^{\prime}}^{2}{\rm
e}^{\nu}\right)^2 }\right].
\end{equation}

As a consequence of the above Eq. (14) we conclude that if charge
vanishes in a charged fluid of embedding class one then the
Schwarzschild interior solution~\cite{Schwarzschild1916} (or
special cases like de Sitter universe or the Einstein universe) or
the Kohler-Chao solution~\cite{Kohlar1965} will be only survived
neutral counterpart unless either the survived spacetime metric is
flat or the charge cannot be zero. Obviously, in the absence of
charge either of two factors on the right hand side of (14) has to
be zero. It can be verified that the vanishing of the first factor
of (14) gives rise to the Kohlar-Chao solution. However, the
vanishing of second factor ultimately provides the Schwarzschild
interior solution.

Let us consider that $m(r)$ is the mass function for electrically
charged fluid sphere and can be given as
\begin{equation}
m(r)={\frac {r}{2}}\,\left[1-{{\rm e}^{-\lambda \left( r \right) }}+{\frac {{q}^{2}}{{r}^{2}}}\right].
\end{equation}

By plugging Eq. (8) and Eq. (14) into Eq. (15), eventually we get
\begin{equation}
m \left(r\right)
=\left[\frac{K\,r\,e^{\nu}\,{\nu^{\prime}}^{2}}{2\,(4+K{\nu^{\prime}}^{2}{\rm
e}^{\nu})}\,+\frac{r\,\left({{K{{\nu}^{\prime}}{{\rm
e}^{\nu}}-2\,r}}\right)\,[\,{\nu}^{\prime}\,(4+K\,{\nu^{\prime}}^{2}{\rm
e}^{\nu})-2\,r\,(2\,{\nu}^{\prime\prime}+{\nu^{\prime}}^{2})]}{4\,\left(4+K{\nu^{\prime}}^{2}{\rm
e}^{\nu}\right)^2 }\right].
\end{equation}

We observe that the expressions for the pressure ($p$), density
($\rho$), electric charge ($q$) and mass ($m$) are dependent on
metric function $\nu$. As a consequence we consider the metric
function $\nu$ to find the spherically symmetric charged fluid
solutions in the following form
\begin{equation}
\nu(r)=n~ln \left(1+A{r}^{2} \right) + ln~B,
\end{equation}
where $n$ is a positive number and $A$ is a constant such that $n
\ge 2$ and $A > 0$.

The above form of the metric potential $\nu(r)$ represents the
same $2\Phi(r)$ as considered by Lake~\cite{Lake2003} in his Eq.
(9) for $A=1/\alpha$ and $B=1$. Therefore, the explanations on Eq.
(17) are mostly the same as in the Ref.~\cite{Lake2003}. The
function $\nu(r)$ is monotone increasing with a regular minimum at
$r = 0$. If we look at the mass function $m(r)$ in Eq. (16), then
it is clear that by using this source function of Eq. (17), the
mass function can easily be evaluated exactly for any $n$. Thus,
the metric function $\nu$ will generate a `class' of solutions
having the physical properties which are expected to be quite
distinct for each value of $n$. It is noted that previously the
solutions for either $n = 1,...,~5$~\cite{Lake1998} or $n \ge
5$~\cite{Maurya2011} solutions were known. With $n = 1,...,~5$,
constitute half of all the previously known physically interesting
solutions in curvature coordinates~\cite{Lake1998} whereas for $N
\ge 5$ solutions are acceptable on physical grounds and even
exhibit a monotonically decreasing subluminal adiabatic sound
speed~\cite{Maurya2011}. It will be also interesting to note that
the above form of $\nu(r)$ is quite different from the function of
Schwarzschild or Kohlar-Chao as one can get a hint from Eq. (14).
Thus, in the present study we expect that each source function
$\nu(r)$ which is a monotone increasing function with a regular
minima at $r = 0$ necessarily provides, via the mass function in
Eq. (16), a static spherically symmetric perfect fluid solution of
Einstein's equations which is regular at $r = 0$.

On the other hand, the metric potential $\lambda (r)$ can be
obtained from Eq. (8) as
\begin{equation}
 \lambda(r)= ln \left[1+K\,A\,B\,n^{2}(Ar^{2})(1+Ar^{2})^{n-2}\right],
\end{equation}
where  $n\geq2$ and $A$, $B$ are positive constants. In Fig. 1 the
behaviour of $\nu (r)$ and $\lambda (r)$ are shown.

\begin{figure}[!htp]\centering
    \includegraphics[width=4.5cm]{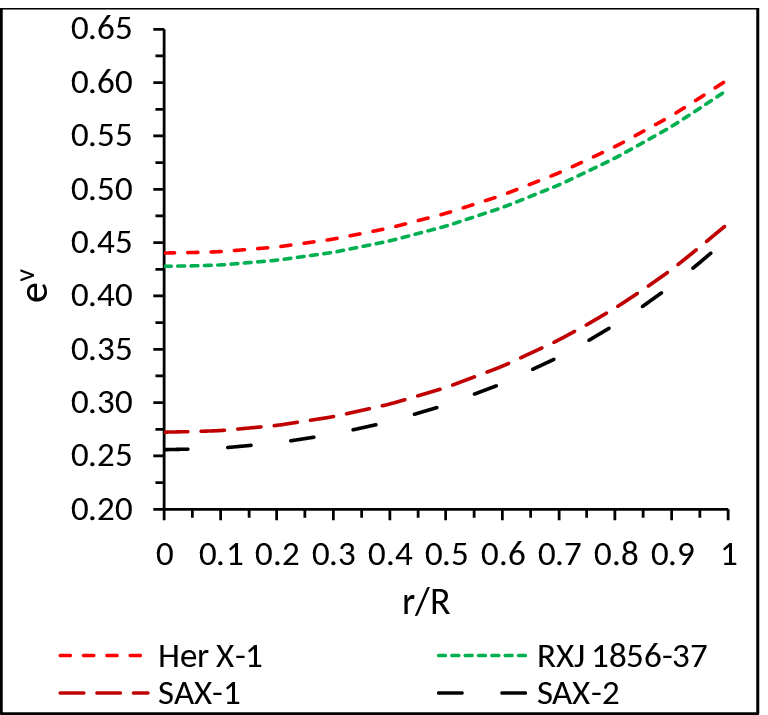}
    \includegraphics[width=4.5cm]{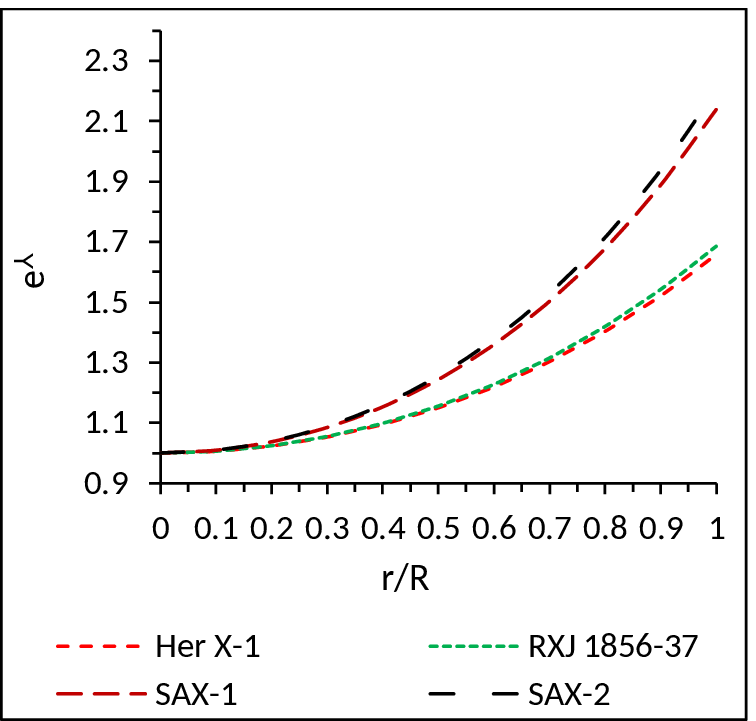}
\caption{Variation of the metric potentials  $e^{\nu}$ (left
panel) and  $e^{\lambda}$ (right panel)  with respect to the
radial coordinate $r/R$ for $n=3.3$. For plotting this figure the
values of the arbitrary constants $A$, $B$ and $K$ are used from
Table 1}
    \label{Fig1}
\end{figure}

The expressions of the electromagnetic mass and the electric
charge are then given by
\begin{equation}
{\frac {2m \left( r \right) }{r}}=A{{r}}^{2}\left[{\frac {n\,f^{2}
\,A{r}^{2}(n - 2) + D f^{n}[-2\,n\, f\,A{r}^{2} +
(2+4\,{A}^{2}{r}^{4}+6\,A{r}^{2}+3\, D
A{r}^{2}\,f^{n})]}{2{\left(1+{A}^{2}{r}^{4}+2\,{{\it Ar}}^{2}+ D
A{r}^{2} f^{n}\right)}^{2}}}\right],
\end{equation}

\begin{equation}
{E}^{2}={A}^{2}{r}^{2}\left[{\frac {n\,f^{2}\,A{r}^{2}(n-2) + D
f^{n}[2\,\left( 1-n \right)\,f +3\,D
f^{n}]}{2{\left(\,1+{A}^{2}{r}^{4}+2\,{{\it Ar}}^{2}+ D A{r}^{2}
f^{n}\right)}^{2}}}\right],
\end{equation}
where $f=\left(1+A{r}^{2}\right)$, $E={\frac {q}{{r}^{2}}}$ and
$D=A\,B\,{n}^{2}K $.

Similarly, the expression for the pressure and the energy density
are given by (Fig. 2)
\begin{equation} 8\,\pi \,p={A\left[\frac {{n}^{2}A{r}^{2}f^{2}-
D\,f^{n} (\,2+2\,A{r}^{2}+D A {r}^{2}f^{n})+2\,n
(1+A{r}^{2})\,p_{1}}{2{\left(1+{A}^{2}{r}^{4}+2\,A{r}^{2}+ D A
{{r}^{2}}f^{n}\right)}^{2}}\right]},
\end{equation}

\begin{equation}
8\,\pi \,\rho=A\left[{\frac {-{n}^{2}A{r}^{2}f^{2}+2 n A
{{r}^{2}}f^{2}(1+3D\,f^{n-1})+Df^{n}\,\rho_{1}}{2{(1+{A}^{2}{r}^{4}+2\,A{r}^{2}+D
A {{r}^{2}}f^{n})}^{2}}}\right],
\end{equation}
where  $p_{1}=[{{2+{A}^{2}{r}^{4}+3\,A{r}^{2}+ D
A{r}^{2}f^{n}}}]$,\,\,$\rho_{1}=[6-4 {A}^{2}{r}^{4}+2 A {r}^{2}+D
A {{r}^{2}}f^{n}]$.

 \begin{figure}[!htp]\centering
    \includegraphics[width=4.5cm]{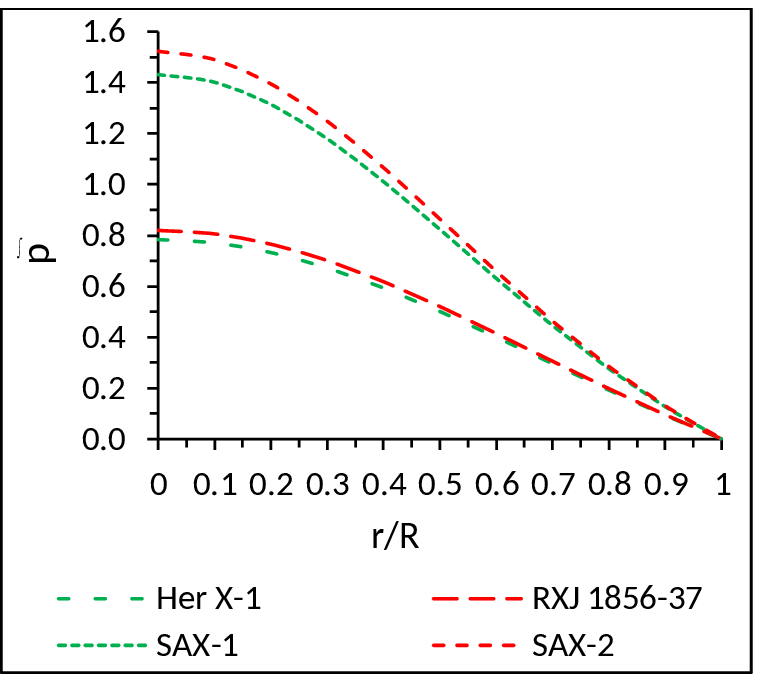}
    \includegraphics[width=4.5cm]{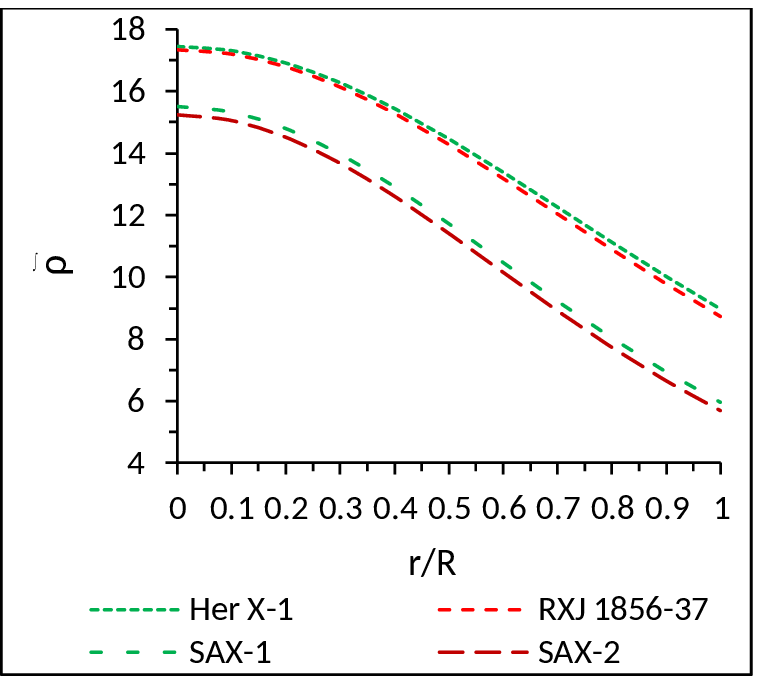}
\caption{Variation of the effective pressure, $\tilde
p=(8\pi/A)p$, and the effective energy density, $\tilde
\rho=(8\pi/A)\rho$, with respect to the radial coordinate $r/R$
for $n=3.3$. For plotting this figure the values of the arbitrary
constants $A$, $B$ and $K$ are used from Table 1}
    \label{Fig.2}
\end{figure}

We suppose that the pressure of the charged fluid sphere is
related with the energy density by a parameter $\omega$ as,
$p=\omega\,\rho$, which is given by (Fig. 3)
\begin{equation}
\omega={A\left[\frac {{n}^{2}A{r}^{2}f^{2}- D\,f^{n}
(\,2+2\,A{r}^{2}+D A {r}^{2}f^{n})+2\,n
(1+A{r}^{2})\,p_{1}}{-{n}^{2}A{r}^{2}f^{2}+2 n A
{{r}^{2}}f^{2}(1+3D\,f^{n-1})+Df^{n}\,\rho_{1}}\right]}.
\end{equation}

\begin{figure}[!htp]\centering
    \includegraphics[width=4.5cm]{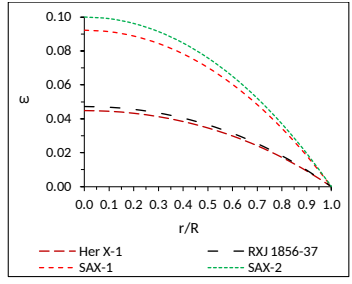}
    \caption{Variation of the parameter $\omega$ with
respect to the radial coordinate ($r/R$). For plotting this figure
the values of the arbitrary constants $A$, $B$ and $K$ are used
from Table 1}
    \label{Fig3}
\end{figure}

We note from Fig. (3) that the ratio $\omega=p/\rho$ is less than
1 throughout inside the star. This obviously implies that the
densities are dominating over the pressures everywhere inside the
star and the underlying fluid distribution is non-exotic in
nature~\cite{Rahaman2010}.

The expressions for the pressure gradients (by taking $ x = A
{r}^{2}$) are given by
\begin{equation}
{\frac {{\it dp}}{{\it dr}}}=-{\frac {2{A}^{2}r}{8\pi
}}\left[{\frac {P_{{1}}+P_{{2}}+P_{{3}}}{2{(1+2\,A
{r}^{2}+A^{2}{r}^{4}+ D A{r}^{2} f ^{n})}^{3}}}\right],
\end{equation}

\begin{equation}
{\frac {{\it d\rho}}{{\it dr}}}=-{\frac {2{A}^{2}r}{8\pi
}}\left[{\frac {\rho_{{1}}+\rho_{{2 }}+\rho_{{3}}}{2{(1+2\,A
{r}^{2}+A^{2}{r}^{4}+ D A{r}^{2} f ^{n})}^{3}}}\right],
\end{equation}
where
\begin{equation} P_{{1}}=2\,D
{n}^{3}{x}^{2}f^{n+1}-{n}^{2}(1+2\,x-2\,{x}^{3}-{x}^{4})+D{n}^{2}x
f^{n} (7+8\,x+{x}^{2}+2\,D x f^{n}),
\end{equation}

\begin{equation}
 P_{{2}}=2\,n f(3+7\,x+{x}^{3}+5\,{x}^{2}) +2\,n D\,f^{n+1}(4+3\,x+{x}^{2}+ D
 x\,f^{n}),
\end{equation}

\begin{equation}
P_{{3}}=- D
f^{n}(6+12\,x+6\,{x}^{2})-{D}^{2}f^{2\,n}(3+4\,x+3{x}^{2}+D
x\,f^{n}),
\end{equation}

\begin{equation}
\rho_{{1}}=2\,D
{n}^{3}{x}^{2}f^{n+1}-{n}^{2}(1+2\,x-2\,{x}^{3}-{x}^{4})+D
{n}^{2}x f^{n}[8\,x+{x}^{2}+7-6\,D xf^{n}],
\end{equation}

\begin{equation}
\rho_{{2}}=-
Df^{n}(22+36\,x+6\,{x}^{2}-8\,{x}^{3})-{D}^{2}f^{2\,n}(11+4\,x+3\,{x}^{2}+{D}^{2}xf^{n}),
\end{equation}

\begin{equation}
\rho_{{3}}=2\,n({1+2\,x-2\,{x}^{3}-{x}^{4}})-2\,n Df^{n}[-6-3\,x+10\,{x}^{2}+7\,{x}^{3}+(5-3{x})\,D x\,f^{n}].
\end{equation}

\section{Matching condition}
For any physically acceptable charged solution, the following
boundary conditions must be satisfied:

\noindent (i) The interior of metric (1) for the charged fluid
distribution join smoothly with the exterior of
Reissner-Nordstr{\"o}m metric

\begin{equation}
ds^{2} =-\left( 1-\frac{2M}{r} +\frac{Q^{2} }{r^{2} } \right)
^{-1} dr^{2} -r^{2} (d\theta ^{2} +\sin ^{2} \theta d\phi ^{2}
)+\left( 1-\frac{2M}{r} +\frac{Q^{2} }{r^{2} } \right) dt^{2},
\end{equation}
at the surface of charged compact stars, whose mass is same as $M$ at $r=R$.

\noindent (ii) The pressure $p$ must be finite and positive at the
centre $r=0$ and it must be zero at the surface $r = R$ of the
charged fluid sphere~\cite{Misner}.

By matching the first and second fundamental forms, the interior
of the metric (2) and the exterior of the metric (32) at the
boundary $r = R$ (the Darmois-Israel condition), we can find the
constants $D$, $B$ and $M$. These are therefore can be obtained as
follows:
\begin{equation}
    \label{19}
    D=\frac{-F^n+(n-1)AR^2\,F^n+nA^2R^4\,F^n+\sqrt{\Psi(R)\,F^{2(n+1)}}}
{AR^2\,F^{2n}},
\end{equation}

\begin{equation}
    \label{20}
    B=(1+AR^2)^{-n}[1+D\,AR^2\,(1+AR^2)^{n-2}]^{-1},
\end{equation}

\begin{equation}
M=\frac{A{{R}}^{3}}{2}\left[{\frac {n\,F^{2}\,A{R}^{2}(n-2) + D
F^{n}[-2\,n\,\,A{R}^{2}\,F + 2\,F\,(2\,F-1)+3\, D
A{R}^{2}\,F^{n}]}{2{\left(F^{2}+ D A{R}^{2}
F^n\right)}^{2}}}\right],
\end{equation}
where $F=(1+AR^2)$, $\Psi(R)=1+2n^2A^2R^4+2\,n\,A\,R^2\,F$.

However, the value of constant $A$  can be determined by assuming
density at the surface of the star i.e. $\rho_{s}$ at $r=R$, so
that we get
\begin{equation}
    \label{23}
A={8\,\pi\,\rho_{s}}\left[\frac{2{(F^2+2\,A{r}^{2}+D A
{{R}^{2}}F^{n})}^{2}}{-{n}^{2}A{R}^{2}F^{2}+2\, n\, A\,
{{R}^{2}}F^{2}(1+3D\,F^{n-1})+D\,F^{n}\,\Omega(R)}\right],
\end{equation}
where $\Omega(R)=[6-4 {A}^{2}{R}^{4}+2 A {R}^{2}+D A
{{R}^{2}}F^{n}]$.

Also the value of constant $K$ can be determined by using the
relation $D=A\,B\,{n}^{2}\,K$ as
 \begin{equation}
    K=\frac{D}{n^2\,A\,B}.
\end{equation}

\begin{table}
\caption{Values of the model parameters $D$,~$A$,~$B$ and $K$ for
different charged compact stars for $n=3.3$} \label{Table1}
\hspace*{-2cm}

{\begin{tabular}{@{}ccccccc@{}} \hline

Compact star  & $D$  &$A$ &$B$ &$K$  \\ candidates & &($cm^{-2}$)
& & $(cm^2)$ \\

\hline Her. X-1 & 5.8164 & $2.2296\times10^{-13}$ & 0.4403 &
5.4406$\times 10^{12} $ \\

\hline RXJ 1856-37 & 5.7794 & $2.8960\times10^{-13}$ & 0.4274 &
4.2877$\times 10^{12} $ \\

\hline SAX J1808.4-3658(SS1)& 5.1682 & $3.5583\times10^{-13}$ &
0.2725 & 4.8944$\times 10^{12} $ \\

\hline SAX J1808.4-3658(SS2)& 5.0772 & $4.7255\times10^{-13}$ &
0.2560 & 3.8540$\times 10^{12} $ \\ \hline
    \end{tabular}}
\end{table}

\section{Physical features of the charged compact star models}
Let us look at the results so far we have obtained in the previous
section. A close observation of the results immediately reveals
the following two distinct features:

\noindent (i) The metric (2) becomes flat and also the expressions
for all the physical parameters, viz. pressure, energy density,
electric charge etc., become zero in all the cases if we take
$A=0$. This feature shows that our solutions represent the
so-called `electromagnetic mass model'~\cite{Lorentz1904}.

\noindent (ii) In this work we have taken the metric function
$\nu(r)$, with $n \geq 2$ and have calculated the data of the
stellar models for $n=3.3$ to $1000$. We come across a very
interesting result that when we increase the value of $n$ at very
large, say more than 100, then the product $nA$ becomes
approximately a constant $C$ (see Table 2). So for the limit $n$
tends to infinity, the present metric potential $\nu(r)$ in Eq.
(17) will convert to the following form: $$\nu(r)=C{r}^{2}+
ln~B,$$ which is the same as considered by Maurya et
al.~\cite{Maurya2015a}. However, the nature of the present models,
at very large value of $n$ i.e. at infinity, can be seen in
Ref.~\cite{Maurya2015a}.

Let us now, besides the above two general features, try to explore
some other physical behaviour of our models.

\subsection{Regularity condition}

\noindent (i) Potentials at the centre $r=0$: From Eqs. (17) and
(18), we observe that the metric potentials at the centre $r=0$
becomes $e^{\lambda(0)}=1$ and $e^{\nu(0)}=B$. This implies that
metric potentials are singularity free and positive at the centre.
However, both are monotonically increasing function (Fig. 1).

\noindent (ii) Pressure at the centre $r=0$: From Eq. (21), one
can obtain $p_0=A\,(2n-D)/8\,\pi$, where $A$ and $D$ are positive
numbers. Hence, the pressure should be positive at the centre and
this implies that $D<2n$.

\noindent (iii) Density at the centre $r=0$: From Eq. (22), we get
the central density $\rho_{0}=(3\,A\,D/8\,\pi)$ which must be
positive at the centre. Since $A$ is positive so $D$ is also
positive due to positivity of $\rho$. We know that
$D=A\,B\,n^{2}\,K$, where $A$, $B$, $n$ all are positive. This
implies that $K$ is also a positive quantity.

\subsection{Casuality and Well behaved condition}
The speed of sound must be less than the speed of light i.e. $0\le
V=\sqrt{dp/d\rho}<1$. However, for well behaved nature of charged
solution, Canuto~\cite{Canuto} argued that the speed of sound
should monotonically decrease outwards for the equation of state
with an ultra-high distribution of matter. Form Fig. 4, one can
observe that the speed of sound is monotonically decreasing
outwards. This implies that our model for charged fluid is well
behaved.

\begin{figure}[!htp]\centering
    \includegraphics[width=4.5cm]{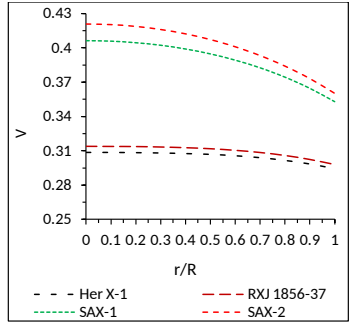}
\caption{Variation of velocity of sound with respect to radial
coordinate r/R for $n=3.3$. For plotting this figure the values of
the arbitrary constants $A$, $B$ and $K$ are used from Table 1}
    \label{Fig.4}
\end{figure}

It can also be observed from Fig. 4 that the velocity of sound
starts decreasing from $n=3.3$ and this clearly indicates that the
solution is physically valid for the values from $n=3.3$ onwards.
However, one thing is then important to know what will happen for
increasing of $n$ towards very large value. It seems possible to
get a reasonable model even when $n$ tends to infinity. This is
because the product of $nA$ becomes approximately constant for
large value of $n$. So if we take $n$ tends to infinity the metric
$\nu$ reduces to the case of Ref.~\cite{Maurya2015a} as discussed
earlier in the introductory part of this Sect. 5.

\subsection{Energy Conditions} For physically valid charged fluid
sphere, the null energy condition (NEC), strong energy condition
(SEC) and weak energy condition (WEC), all must satisfy
simultaneously at all the interior points of the star. Therefore,
in our model the following inequalities should hold good:\\

\hspace*{-0.55cm} NEC: $\rho+\frac{{E}^{2}}{8\pi}\geq0$,
\hspace*{0.55cm} WEC: $\rho-p+\frac{{E}^{2}}{4\pi}\geq0$,
\hspace*{0.55cm} SEC:  $\rho-3p+\frac{{E}^{2}}{2\pi}\geq0$.

\begin{figure}[!htp]\centering
    \includegraphics[width=4.5cm]{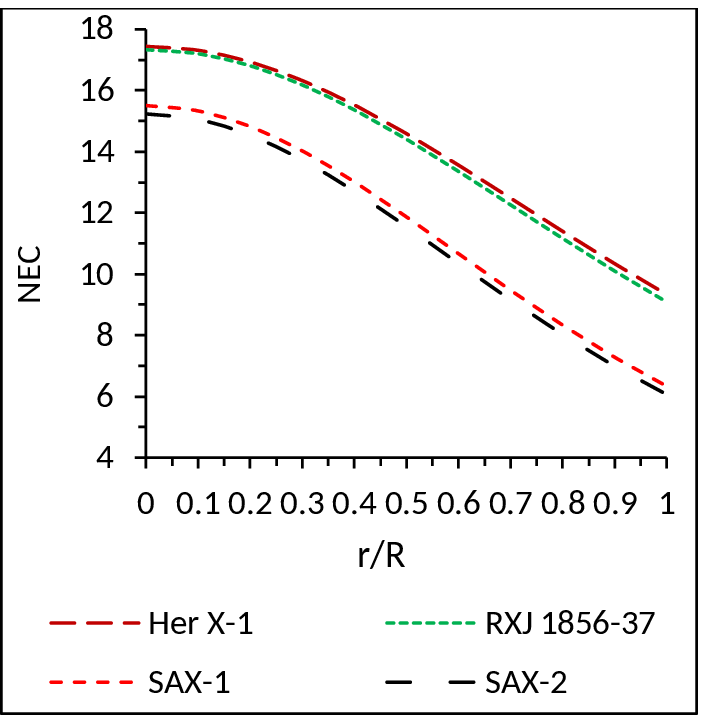}
    \includegraphics[width=4.5cm]{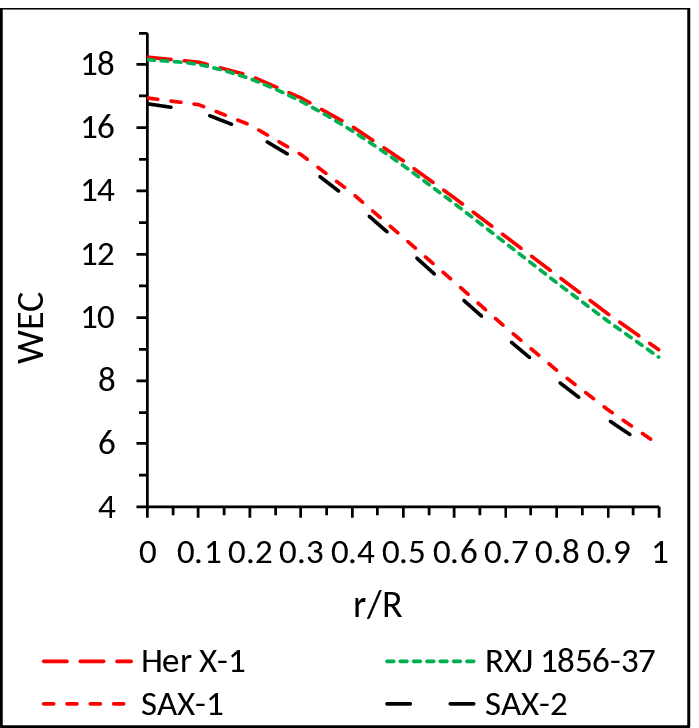}
    \includegraphics[width=4.5cm]{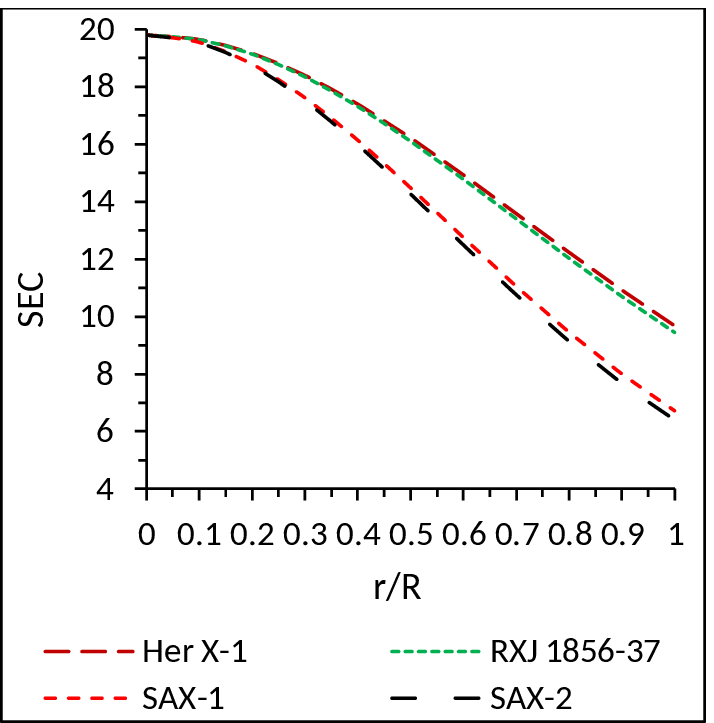}
\caption{Variation of energy conditions with respect to radial
coordinate r/R for $n=3.3$. For plotting this figure the values of
the arbitrary constants $A$, $B$ and $K$ are used from Table 1}
    \label{Fig.5}
\end{figure}

In Fig. 5 we have shown the energy conditions which are as par
physical requirement.

\subsection{Generalized TOV equation} The generalized
Tolman-Oppenheimer-Volkoff (TOV)
equation~\cite{Tolman1939,Oppenheimer1939}
\begin{equation}
-\frac{M_G(\rho+p_r)}{r^2}e^{\frac{\lambda-\nu}{2}}-\frac{dp}{dr}+
\sigma \frac{q}{r^2}e^{\frac{\lambda}{2}} =0,
\end{equation}
where $M_G$ is the effective gravitational mass given by
\begin{equation}
M_G(r)=\frac{1}{2}r^2 \nu^{\prime}e^{(\nu - \lambda)/2}.
\end{equation}

Eq. (38) describes the equilibrium condition for a charged perfect
fluid subject to the sum total interaction between the
gravitational $(F_g)$, hydrostatic $(F_h)$ and electric $(F_e)$,
so that one should get
\begin{equation}
F_g+F_h+F_e=0,
\end{equation}
where
\begin{equation}
F_g=-\frac{1}{2}\left( \rho+p\right) {{\nu}^{\prime}}=-{\frac
{n{A}^{2}r}{8\pi }}\left[{\frac {2\, D f^{n+1} \left(
1-A{r}^{2}+2\,nA{r}^{2} \right) +2\,n f^{3}}{2\, f
{(1+2\,A{r}^{2}+{A}^{2}{r}^{4}+ D A{r}^{2} f^{n})}^{2}}}\right],
\end{equation}

\begin{equation}
F_h=-\frac{dp}{dr},
\end{equation}

\begin{equation}
F_e={\frac {{A}^{2}r}{4\pi }}\left[{\frac {F_{{{\it e1}}}+F_{{{\it
e2}}}+F_{{{\it e3}}}+F_{{{\it e4}}}+F_{{{\it
e5}}}}{2\,{(1+2\,A{r}^{2}+{A}^{2}{r}^{4}+ D A{r}^{2}
f^{n})}^{3}}}\right],
\end{equation}

\begin{equation}
F_{e1}=-2\, D {n}^{3}{A}^{2}{r}^{4}
f^{n+1}+{n}^{2}(3+10\,A{r}^{2}+12\,{A}^{2}{r}^{4}+6\,{A}^{3}{r}^{6}+{A}^{4}{r}^{8}),
\end{equation}

\begin{equation}
F_{e2}={n}^{2} D
A{r}^{2}f^{n}(-1+4\,A{r}^{2}+5\,{A}^{2}{r}^{4}+2\,D A{r}^{2}
f^{n}),
\end{equation}

\begin{equation}
F_{{{\it e3}}}= Df^{n}[6+12\,A{r}^{2}+6\,{A}^{2}{r}^{2}+ D f^{n}
(3+4\,A{r}^{2}+3\,{A}^{2}{r}^{4}+ D A{r}^{2} f^{n})],
\end{equation}

\begin{equation}
F_{{{\it e4}}}=
2\,n[3+10\,A{r}^{2}+12\,{A}^{2}{r}^{4}+6\,{A}^{3}{r}^{6}+{A}^{4}{r}^{8}],
\end{equation}

\begin{equation}
F_{{{\it e5}}}=2\,n  D f^{n}[\,3+2\,{A}^{3}{r}^{6}+6\,A{r}^{2}+5\,{A}^{2}{r}^{4}+2\,D{A}^{2}{r}^{4} f^{n}].
\end{equation}

\begin{figure}[!h]\centering
    \includegraphics[width=4.5cm]{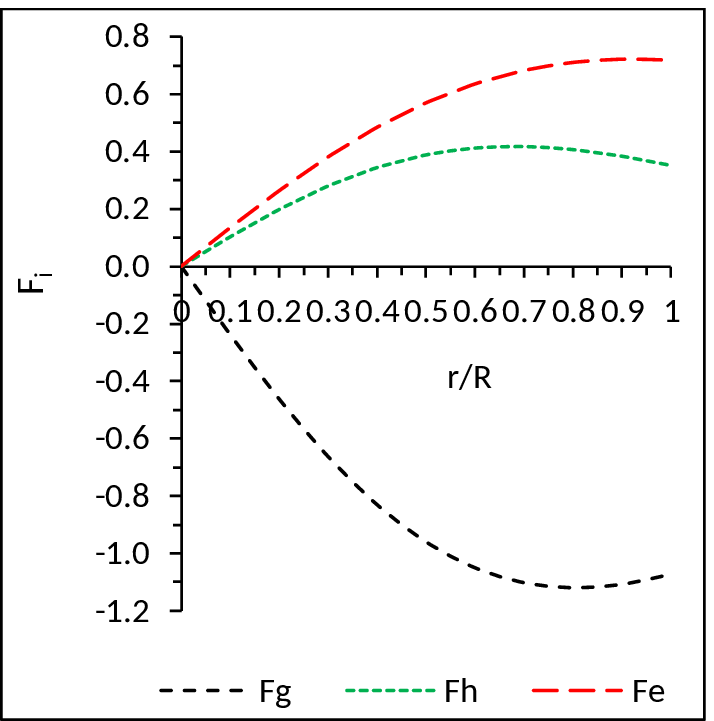}
    \includegraphics[width=4.5cm]{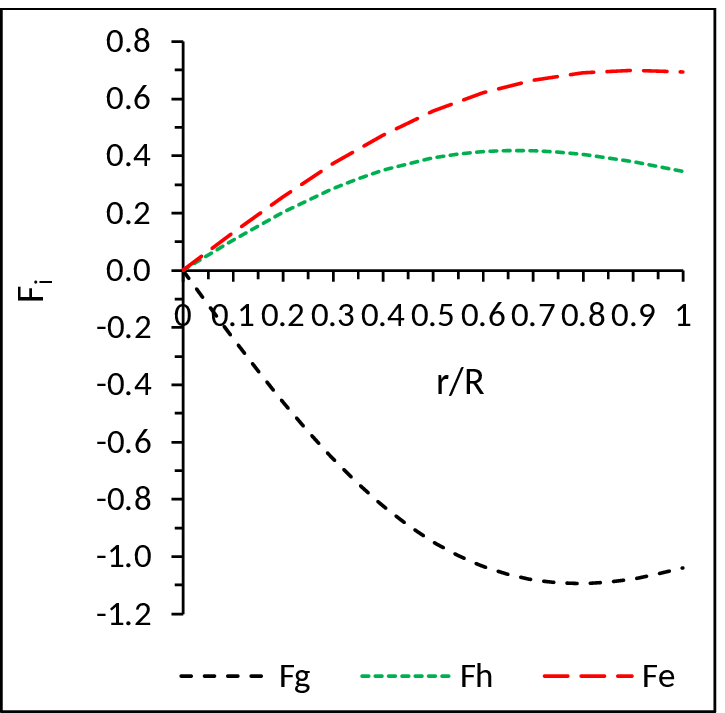}
     \includegraphics[width=4.5cm]{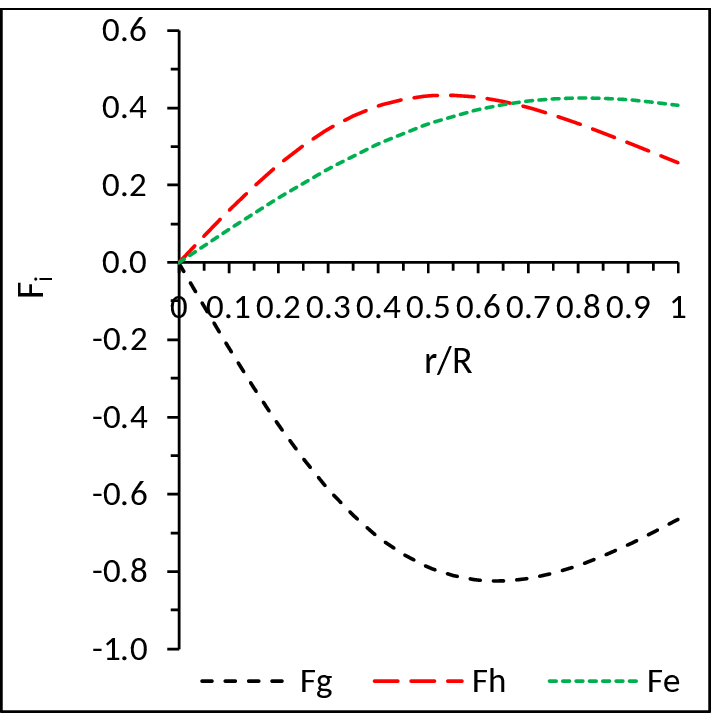}
     \includegraphics[width=4.5cm]{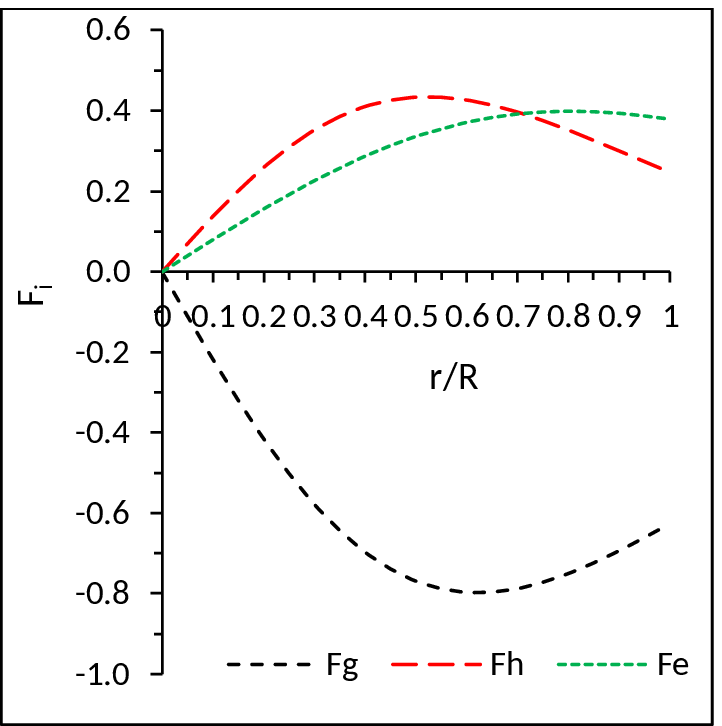}
\caption{Variation of different forces with respect to the radial
coordinate $r/R$ for $n=3.3$. (i) Her X-1 (top left), (ii) RXJ
1856-37 (top right), (iii)SAX J1808.4- 3658(SS1), (iv) SAX
J1808.4-3658(SS2). For plotting this figure the values of the
arbitrary constants $A$, $B$ and $K$ are used from Table 1}
    \label{Fig.6}
\end{figure}

From the plot for TOV equation in Fig. 6 it can be observed that
the system is in static equilibrium. The sum of all the forces,
like gravitational, hydrostatic and electric forces, are zero. It
is interesting to note from Fig. 6 that the gravitational force is
counter balanced by the joint action of hydrostatic and electric
forces.

\subsection{Effective mass-radius relation} For physically valid
models, the ratio of the mass to the radius of a compact star
models can not be arbitrarily large. Buchdahl \cite{Buchdahl1959}
has imposed a stringent restriction on the mass-to-radius ratio
that for the perfect fluid model it should be $2M/R < 8/9$.
However, B{\"o}hmer and Harko \cite{Boehmer2007} have given the
generalized expression of lower bound for charged compact object
as follows:
\begin{equation}
\frac{3{{Q}^{2}}}{2{{R}^{2}}}\frac{\left(1+\frac{{{Q}^{2}}}{18{{R}^{2}}}\right)}{\left(1+\frac{{{Q}^{2}}}{12{{R}^{2}}}\right)}\leq
\frac{2M}{R}.
\end{equation}

The upper bound of the mass for charged fluid sphere was
generalized  by Andr{\'e}asson \cite{Andreasson} and proved that
\begin{equation}
\sqrt{M}\leq\frac{\sqrt{R}}{3}+\sqrt{\frac{R}{9}+\frac{{{Q}^{2}}}{3R}}.
\end{equation}

We, therefore, conclude from the above two conditions that $2M/R$
must satisfy the following inequality
\begin{equation}
\frac{3{{Q}^{2}}}{2{{R}^{2}}}\frac{\left(1+\frac{{{Q}^{2}}}{18{{R}^{2}}}\right)}{\left(1+\frac{{{Q}^{2}}}{12{{R}^{2}}}\right)}\leq
\frac{2M}{R}\leq
\frac{2}{R}\left(\frac{\sqrt{R}}{3}+\sqrt{\frac{R}{9}+\frac{{{Q}^{2}}}{3R}}\right)^{2}.
\end{equation}

In this model, the effective gravitational mass has the following
form
\begin{equation}
M_{{{\it eff}}} =4{\pi}\int _{0}^{R}\! \left( \rho+{\frac
{{E}^{2}}{8\pi }} \right) {r}^{2}{dr}=\frac{1}{2}R[1-{{\rm
e}^{-\lambda \left( R \right) }}],
\end{equation}
which can finally be expressed as
\begin{equation}
M_{{{\it eff}}}=\frac{1}{2}R[{{{\frac { D A{R}^{2} \left(
1+A{R}^{2} \right) ^{n-2}}{1+ D A{R}^{2} \left( 1+A{R}^{2}\right)
^{n-2}}}}}].
\end{equation}

\subsection{Surface red-shift} We define the compactification
factor as
\begin{equation}
u=\frac{M_{eff}}{R}=\frac{1}{2}[1-{{\rm e}^{-\lambda \left( R
\right)}}]=\frac{1}{2}[{{{\frac { D A{R}^{2} \left(1+A{R}^{2}
\right)^{n-2}}{1+ D A{R}^{2} \left( 1+A{R}^{2}\right)^{n-2}}}}}].
\end{equation}

The surface redshift corresponding to the above compactness factor
$u$ is obtained as
\begin{equation}
Z = {\left(1-2u\right)}^{-1/2}-1=\sqrt{1+DA{{R}^{2}}\left(
1+A{R}^{2}\right) ^{n-2}}-1.
\end{equation}

\begin{figure}[!htp]\centering
    \includegraphics[width=4.5cm]{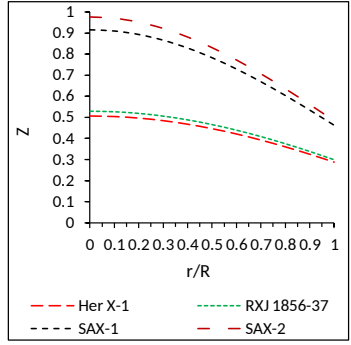}
\caption{Variation of redshift with respect to the radial
coordinate $r/R$ for $n=3.3$. For plotting this figure the values
of the arbitrary constants $A$, $B$ and $K$ are used from Table 1}
    \label{Fig.7}
\end{figure}

\begin{table}
\caption{Numerical values of the product of $n$ and $A$ i.e. $nA$
for different charged compact star models} \label{Table 2}
\hspace*{-2cm}

{\begin{tabular}{@{}ccccccc@{}}

\hline

& $n=3.3$ & $n=10$ & $n=100$ & $n=1000$ \\

\hline Compact stars & $nA$ & $nA$ & $nA$ &$nA$\\

\hline Her. X-1 & 7.36x${10}^{-13}$ & 7.32x${10}^{-13}$ &
7.29x${10}^{-13}$ & 7.29x${10}^{-13}$ \\

\hline RXJ 1856-37 & 9.56x${10}^{-13}$ & 9.47x${10}^{-13}$ &
9.43x${10}^{-13}$ & 9.44x${10}^{-13}$\\

\hline SAX J1808.4- 3658(SS2) & 15.59x${10}^{-13}$ &
15.25x${10}^{-13}$ & 15.10x${10}^{-13}$  & 15.08x${10}^{-13}$\\

\hline SAX J1808.4- 3658(SS1) & 11.74x${10}^{-13}$ &
11.50x${10}^{-13}$ & 11.40x${10}^{-13}$  & 11.39x${10}^{-13}$\\

\hline
\end{tabular}}
\end{table}

\begin{table}
\centering \caption{Numerical values of physical parameters
$M\left(M_\odot\right)$, $R~(km)$ and $A{R}^{2}$ for different
values of $n$} \label{Table 3}

{\begin{tabular}{@{}ccccccc@{}}

\hline

& & & $ n=3.3$ & $n=10$ & $n=100$ & $n=1000$ \\

\hline

Compact star & $M~(M_\odot)$ & $R~(km)$ & $A{R}^{2}$ & $A{R}^{2}$
& $A{R}^{2}$ & $A{R}^{2}$ \\

\hline Her. X-1 & 0.98 & 6.7  & 0.1000 & 0.03272 & 0.003260 &
0.0003258 \\

\hline RXJ 1856-37 & 0.9048 & 6.003  & 0.10437 & 0.03414 &
0.003400 & 0.0003400 \\

\hline SAX J1808.4- 3658(SS2) & 1.3232 & 6.33  & 0.1893 & 0.06108
& 0.006048 & 0.0006042 \\

\hline SAX J1808.4- 3658(SS1) & 1.435 & 7.07  & 0.1779 & 0.05750 &
0.005700 & 0.0005695 \\

\hline
\end{tabular}}
\end{table}

In Table 3 we have shown $AR^2$ which are very required as all the
equations are dependent on $AR^2$, specially Eq. (35). As we know
that for each different star the ratio $M/R$ is fixed, so for this
purpose we suppose the value of $AR^2$ to determine the ratio
$M/R$ from Eq. (35). The feature of $Z$ is shown in Fig. 7.

\begin{table}
\centering \caption{The energy densities and central pressure for
different charged compact star candidates for $n=3.3$}
\label{Table4}

{\begin{tabular}{@{}ccccccc@{}}

\hline

Compact star & Central density & Surface density & Central
pressure & $2M/R$\\  & ($gm/cm^{3}$) & ($gm/cm^{3}$) &
($dyne/cm^{2}$)
\\\hline

Her. X-1 & 2.0892$\times{10}^{15}$ & 1.0742$\times{10}^{15}$ &
8.4453$\times{10}^{34}$ & 0.432&
\\ \hline

RXJ 1856-37 & 2.6964$\times{10}^{15}$  & 1.3588$\times{10}^{15}$ &
1.1487$\times{10}^{35}$ & 0.444\\ \hline

SAX J1808.4-3658(SS2) & 3.8651$\times{10}^{15}$ &
1.4409$\times{10}^{15}$ & 3.4786$\times{10}^{35}$ & 0.616\\ \hline

SAX J1808.4-3658(SS1) & 2.9626$\times{10}^{15}$ &
1.1407$\times{10}^{15}$ & 2.4628$\times{10}^{35}$ & 0.598\\

\hline
\end{tabular}}
\end{table}

\subsection{Electric charge} The amount of charge at the centre and
boundary for different stars are given in Table 5. Also, from Fig.
8, it is clear that the charge profile is minimum at the centre
and monotonically increasing away from the centre, however it
acquires the maximum value at the boundary of the stars. To
convert the amount of charge in Coulomb, every value should be
multiplied by a factor $1.1659 \times10^{20}$ in the Table 5.

\begin{table}
\caption{The electric charge for different compact stars in the
relativistic unit (km)} \label{Table1}

{\begin{tabular}{@{}ccccccc@{}} \hline

$r/a$ & Her. X-1 & RXJ 1856-37 & SAX J1808.4-3658(SS1) & SAX
J1808.4-3658(SS2) \\

\hline 0.0 & 0.0 &  0.0 & 0.0 & 0.0 \\

\hline 0.2 & 0.0126 &  0.0117 & 0.0189 & 0.0174\\

\hline 0.4 & 0.098 &  0.0905 & 0.1466 & 0.135\\

\hline 0.6 & 0.3144 & 0.2902 & 0.468 & 0.4314\\

\hline 0.8 & 0.6971 & 0.6426 & 1.0253 & 0.9453\\

\hline 1.0 & 1.2564 & 1.1562  & 1.8141 & 1.6706  \\

\hline
\end{tabular}}
\end{table}

\begin{figure}[!htp]\centering
\includegraphics[width=4.5cm]{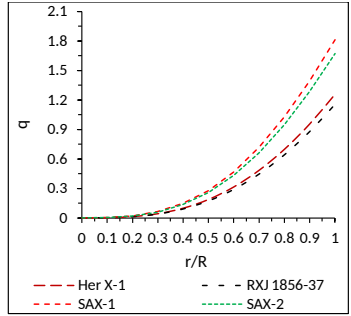}
\caption{Variation of the electric charge ($q$) with respect to
the radial coordinate $r/R$ for $n=3.3$. For plotting this figure
the values of the arbitrary constants $A$, $B$ and $K$ are used
from Table 1}
    \label{Fig8}
\end{figure}

\section{Conclusion}
We have investigated for a new stellar model with spherically
symmetric matter distribution under the Einstein-Maxwell
spacetime. It is observed that the model represents a compact star
of embedding class one. The solutions obtain here are general in
their nature having the following two specific features:

(i) The metric becomes flat and also the expressions for the
pressure, energy density and electric charge become zero in all
the cases if we consider the constant $A=0$, which shows that our
solutions represent the so-called `electromagnetic mass
model'~\cite{Lorentz1904}.

(ii) The metric function $\nu(r)$, for the limit $n$ tends to
infinity, converts to $\nu(r)=C{r}^{2}+ ln~B$, which is the same
as considered by Maurya et al.~\cite{Maurya2015a}.

We have also studied several physical aspects of the model and
find that all the features are acceptable within the expected
demand of the contemporary theoretical works and observational
evidences. Some salient features of these physical behaviour of
our models are as follows:

(1) Regularity condition: We have discussed the situations in the
following cases:

(i) Potentials at the centre $r=0$: From Eqs. (17) and (18), we
observe that the metric potentials at the centre $r=0$ becomes
$e^{\lambda(0)}=1$ and $e^{\nu(0)}=B$. This implies that metric
potentials are singularity free and positive at the centre.
However, both are monotonically increasing function (Fig. 1).

(ii) Pressure at the centre $r=0$: From Eq. (21), one can obtain
$p_0=A\,(2n-D)/8\,\pi$, where $A$ and $D$ are positive. The
pressure should be positive at the centre and this implies that
$D<2n$.

(iii) Density at the centre $r=0$: From Eq. (22), we get the
central density $\rho_{0}=(3\,A\,D/8\,\pi)$ which must be positive
at the centre. Since $A$ is positive then $D$ is also positive due
to positivity of $\rho$. We know that $D=A\,B\,n^{2}\,K$, where
$A$, $B$, $n$ are all positive. This implies that $K$ is also
positive.

(2) Casuality and Well behaved condition: The speed of sound as
suggested by Canuto~\cite{Canuto} satisfies in the presented
compact star model as is evident from Fig. 4.

It can be observed from Fig. 4 that the velocity of sound starts
decreasing from $n=3.3$ and this clearly indicates that the
solution is well behaved from $n=3.3$ onwards and it seems
possible to get a reasonable model even when $n$ tends to
infinity.

(3) Energy Conditions: In our model all the energy conditions,
viz. NEC, SEC and WEC, satisfy simultaneously at all the interior
points of the star.

(4) Generalized TOV equation: The generalized
Tolman-Oppenheimer-Volkoff (TOV)
equation~\cite{Tolman1939,Oppenheimer1939} satisfies here and
indicates that the model is in static equilibrium under the
interaction between the gravitational, hydrostatic and electric
forces.

(5) Effective mass-radius relation: We have verified that the
Buchdahl \cite{Buchdahl1959} condition $2M/R < 8/9$ satisfies in
our model within the stipulated range as can be observed from the
Table 4.

(6) Surface red-shift: The surface redshift in the present model
is found to be satisfactory as can be seen from Fig. 7.

(7) Electric charge: The amount of charge at the centre and
boundary for different stars can be found from Table 5. Fig. 8
depicts that the charge is minimum at the centre and monotonically
increasing away from the centre, however it acquires the maximum
value at the boundary of the stars.

As a final comment, however, it is to be justified to consider
several other aspects of embedding class 1 metric and further
investigations on the corresponding model for compact stars as far
as ultra-modern observational evidences are concerned.

\section*{Acknowledgments}
SKM acknowledges support from the authority of University of
Nizwa, Nizwa, Sultanate of Oman. Also SR is thankful to the
authority of The Institute of Mathematical Sciences, Chennai,
India for providing Associateship under which a part of this work
was carried out there.


\begin{thebibliography}{99}


\bibitem{Lorentz1904} H.A. Lorentz, Proc. Acad. Sci. Amsterdam 6 (1904) (Reprinted in:
Einstein et al., ``The Principle of Relativity'', Dover, INC, p.
24, 1952)

\bibitem{Maurya2015a} S.K. Maurya, Y.K. Gupta, S. Ray, S. Roy Chowdhury, Eur. Phys. J. C {\bf 75} 389 (2015)

\bibitem{Friedmann1922} A. Friedmann, Zeit. Physik {\bf 10}, 377 (1922)

\bibitem{Robertson1933} H.P. Robertson, Rev. Mod. Phys. {\bf 5}, 62 (1933)

\bibitem{Lemaitre1933} G. Lemaître, Annal. Soc. Sci. Brux. {\bf 53}, 51 (1933)

\bibitem{Kuzeev1980} R.R. Kuzeev, Gravit. Teor. Otnosit. \textbf{16}, 93 (1980)

\bibitem{Schwarzschild1916} K. Schwarzschild, Phys.-Math. Klasse, 189 (1916)

\bibitem{Kohlar1965} M. Kohler, K.L. Chao, Z. Naturforsch. Ser. {\bf A 20}, 1537 (1965)

\bibitem{Wheeler1962} J.A. Wheeler, Geometrodynamics (p. 25, Academic, New York, 1962)

\bibitem{Feynman1964} R.P. Feynman, R.B. Leighton and M. Sands,
"The Feynman Lectures on Physics", Addison-Wesley, Palo Alto, Vol.
II, Chap. 28 (1964)

\bibitem{Florides1974} P.S. Florides, Proc. R. Soc. (Lond.) Ser. A \textbf{337}, 529 (1974)

\bibitem{Tiwari1984} R.N. Tiwari, J.R. Rao, R.R. Kanakamedala, Phys. Rev.
D {\bf 30}, 489 (1984).

\bibitem{Gautreau1985} R. Gautreau, Phys. Rev. D {\bf 31}, 1860 (1985).

\bibitem{Gron1985} {\O}. Gr{\o}n, Phys. Rev. D {\bf 31}, 2129 (1985).

\bibitem{Gron1986} $\phi$. Gr$\phi$n, Gen. Relativ. Gravit. \textbf{18}, 591 (1986)

\bibitem{Cooperstock1989} F.I. Cooperstock, N. Rosen, Int. J. Theor. Phys. \textbf{28}, 423 (1989)

\bibitem{Tiwari1991a} R.N. Tiwari, J.R. Rao, S. Ray, Astrophys. Space Sci.
{\bf 178}, 119 (1991)

\bibitem{Tiwari1991b} R.N. Tiwari, S. Ray, Astrophys. Space Sci.
{\bf 180}, 143 (1991)

\bibitem{Tiwari1991c} R.N. Tiwari, S. Ray, Astrophys. Space Sci.
{\bf 182}, 105 (1991)

\bibitem{Ray1993} S. Ray, D. Ray, R.N. Tiwari, Astrophys. Space Sci. \textbf{199}, 333 (1993)

\bibitem{Ray1997} R.N. Tiwari, S. Ray, Gen. Relativ. Gravit. \textbf{29}, 683 (1997)

\bibitem{Wilzchek1999} F. Wilczek, Phys. Today {\bf 52}, 11
(1999)

\bibitem{Ray2002a} S. Ray, Astrophys. Space Sci. \textbf{280}, 345 (2002)

\bibitem{Ray2002b} S. Ray, B. Das, Astrophys. Space Sci. {\bf 282}, 635 (2002)

\bibitem{Ray2004a} S. Ray, B. Das, Mon. Not. R. Astron. Soc. {\bf 349}, 1331 (2004)

\bibitem{Ray2004b} S. Ray, S Bhadra, Phys. Lett. A {\bf 322}, 150
(2004)

\bibitem{Ray2006} S. Ray, Int. J. Mod. Phys. D {\bf 15} 917 (2006)

\bibitem{Ray2007b} S. Ray, B. Das, F. Rahaman, S. Ray, Int. J. Mod. Phys. D {\bf 16}, 1745 (2007)

\bibitem{Mehra1980} A.L. Mehra, Gen. Relativ. Gravit. \textbf{12}, 187 (1980)

\bibitem{Kuchowicz1968} B. Kuchowicz, Acta Phys. Pol. \textbf{33}, 541 (1968)

\bibitem{Krori1975} K.D. Krori, J. Barua, J. Phys. A \textbf{8}, 508 (1975)

\bibitem{Karmarkar} K.R. Karmarkar, Proc. Ind. Acad. Sci.{\bf A 27}, 56 (1948)

\bibitem{Pandey1982} S.N. Pandey, S.P. Sharma, Gen. Relativ. Gravit. {\bf
14}, 113 (1982)

\bibitem{Lake2003} K. Lake, Phys. Rev. D {\bf 67}, 104015 (2003)

\bibitem{Lake1998} M.S.R. Delgaty, K. Lake, Comput. Phys. Commun. {\bf 115},
395 (1998)

\bibitem{Maurya2011} S.K. Maurya, Y.K. Gupta, Astrophys Space Sci. {\bf 334}, 301 (2011)

\bibitem{Rahaman2010} F. Rahaman, S.A.K. Jafry, K. Chakraborty, Phys. Rev. D {\bf 82}, 104055
(2010)

\bibitem{Misner} C.W. Misner, D.H. Sharp: Phys. Rev. {\bf B 136}, 571 (1964)

\bibitem{Canuto} V. Canuto, In Solvay Conf. on Astrophysics and Gravitation, Brussels (1973)

\bibitem{Tolman1939} R.C. Tolman, Phys. Rev. {\bf 55}, 364 (1939)

\bibitem{Oppenheimer1939} J.R. Oppenheimer, G.M. Volkoff, Phys. Rev. {\bf 55} 374 (1939)

\bibitem{A} S.K. Maurya, Y.K. Gupta, S. Ray, arXiv: {\bf 1502.01915}, [gr-qc] (2015)

\bibitem{Buchdahl1959} H.A. Buchdahl, Phys. Rev. {\bf 116} 1027 (1959).

\bibitem{Boehmer2007} C.G. B{\"o}hmer and T. Harko, Gen. Relativ. Gravit. {\bf 39} 757 (2007).

\bibitem{Andreasson} H. Andr{\'e}asson, Commun. Math. Phys. {\bf 288}, 715 (2009)

\end{thebibliography}
\end{document}